\documentclass[twocolumn,showpacs,showkeys]{revtex4}
\usepackage{graphicx}
\usepackage{bm}
\usepackage{color}
\usepackage{amsmath}
\usepackage{natbib}

\begin{document}

\title{Twisted behavior of dipolar BECs: Dipole-dipole interaction beyond the self-consistent field approximation and exchange electric dipole interaction}

\author{Pavel A. Andreev}
\email{andreevpa@physics.msu.ru}
\affiliation{Faculty of physics, M. V. Lomonosov Moscow State University, Moscow, Russia.}

\date{\today}

\begin{abstract}
Dipole-dipole interaction is a long-range interaction, hence we could expect that the self-consistent field approximation might be applied. In most cases it is correct, but dipolar BECs reveal a surprise. Structure of the self-consistent field term requires that interacting particles are in different quantum states, while in BECs all particles in a single quantum state. This fact requires to consider the two-particle polarisation, which describes dipole-dipole interaction, in more details. We present this consideration and show an astonishing result that the two-particle quantum correlation in dipolar BECs reveals in the same form as the self-consistent field term.
\end{abstract}


\maketitle

\section{Introduction}

Classic papers on dipolar BECs \cite{Santos PRL 00}, \cite{Goral PRA 00}, \cite{Yi PRA 00}, suggesting a generalisation of the Gross-Pitaevskii equation, do not describe approximation for dipole-dipole interaction in BECs. They just note that long-range and anisotropic nature of dipole-dipole interaction is highly interesting. One of the following papers \cite{Yi PRA 01} discusses justification of the model in terms of scattering that does not fully satisfy picture of the long-range interaction developed in plasmas physics in 30-th of XX century \cite{Vlasov UFN}.

After many applications the physical picture giving understanding of the approximation suggested in Refs. \cite{Santos PRL 00}, \cite{Goral PRA 00}, \cite{Yi PRA 00} has not appeared. There are a lot of reviews on dipolar BECs, see for instance \cite{Lahaye RPP 09} and \cite{Baranov CR 12}.

Recently we have derived models of dipolar BECs in different approximations: align dipoles \cite{Andreev MPL 13}, \cite{Andreev EPJ D 14}, dipoles with evolution of dipole directions (see \cite{Andreev EPJ D 13}, \cite{Andreev TransDipBEC12} for BECs, and \cite{Andreev RPJ 13} for ultracold fermions). All these papers are based on earlier papers on dynamics of dipolar charged particles at finite temperatures \cite{Maksimov TMP 01}, \cite{Andreev RPJ 07}, \cite{Andreev PRB 11}.

In Refs. \cite{Andreev MPL 13}-\cite{Andreev PRB 11} the dipole-dipole interaction is considered as a long-range interaction. Consequently the self-consistent field approximation was applied. Formally the self-consistent field approximation means that two-particle hydrodynamic functions appearing in the force field are represented as product of corresponding one-particle functions, for instance the two-particle concentration $n_2(\textbf{r},\textbf{r}',t)$ is replaced by the product of concentrations $n(\textbf{r},t) n(\textbf{r}',t)$.

We will show below that the self-consistent field approximation requires that interacting particles are in different quantum states. Consequently this approximation works well for systems with thermally distributed particles, and ultracold fermions distributed over the large number low laying quantum states due to the Pauli principle. However the self-consistent field approximation can not be applied to BECs, where all particles are located in a single quantum state.
Same conclusion is correct for magnetic dipolar BECs.

Similar picture appears for the Coulomb interaction of Cooper pair of electrons in superconductors. Interaction of charges arises in the form similar to the self-consistent field, but its nature lays in the exchange Coulomb interaction of bosons located in the BEC state, and the self-consistent field equals to zero.

A big step towards consideration of quantum Bose and Fermi gases was made in Ref. \cite{Baillie PRA 12}, where authors developed a kinetic model for finite temperature gas. This model appears to have complicate structure. However this paper does not contain exhaustive analysis of the zero temperature limit, which is essential for physics of ultracold gases. Hence authors of Ref. \cite{Baillie PRA 12} did not uncover unusual behavior of the model of dipolar BECs described in our paper.

In  this paper we discuss background of minimal coupling model for dipolar BECs, but we can point out on resent generalisations of this model. For instance in Ref. \cite{Wang NJP 08}, author considered the dipole-dipole interaction as a scattering process and suggested a way of generalisation of the standard model \cite{Santos PRL 00}, \cite{Goral PRA 00}, \cite{Yi PRA 00} going beyond the first Born approximation. The presence of particles in states with non-minimal
energy (quantum fluctuations) due to interparticle interactions, and their influence on the properties of BECs, were considered in Ref. \cite{Lima PRA 12}, leading to an apparatus for generalising the GP equation for dipolar, fully polarised, BECs. Following by Popov's steps, authors of Ref. \cite{Natu_Wilson PRA 13} considered the Landau damping in a collisionless dipolar Bose gas. Dipolar BECs with the Rashba spin-orbit interaction was considered in Ref. \cite{Wilson PRL 13}.

The standard model also find a lot of new applications presented in Refs. \cite{Natu PRA 14}-\cite{Kumar EPJD 14}. Speaking about application of the standard model of dipolar BECs we should mention that we explicitly apply the full potential of the electric dipole interaction (see formula (\ref{Twb dBEC dd int Green funct}) below, or papers \cite{Andreev MPL 13}, \cite{Andreev EPJ D 14}, \cite{Andreev TransDipBEC12}), whereas the standard model contains the shorted potential with no delta function term.

Presented in this paper conclusions about the nature of the dipole-dipole interaction do not depend on the explicit form of the dipole-dipole potential. However, the non-integral form of equations and explicit form of equations of field are directly related to the explicit form of the potential of dipole-dipole interaction.

Two dimensional dipolar BECs were considered in Refs. \cite{Lu_Shlyapnikov arX_14}, \cite{Boudjemaa PRA 13}. It seems that problem of presence of the delta function in the dipole-dipole interaction potential was ignored there. Hence we briefly debate this problem below in this paper.

This paper is organized as follows. In Sec. II approximations for model of dipolar BECs with aligned dipoles is examined. In Sec. III we examine the model of dipolar BECs with the dipole direction evolution. In Sec. IV we summarize results of the model analysis presented in Sec. II and III.

\section{Model}

At derivation of equations describing collective evolution of quantum systems from the many-particle Schrodinger equation we obtain the continuity and Euler equations:
\begin{equation}\label{Twb dBEC cont eq from GP}
\partial_{t}n+\nabla\cdot(n\textbf{v})=0,\end{equation}
and
$$mn(\partial_{t}+\textbf{v}\cdot\nabla)\textbf{v}-\frac{\hbar^{2}}{4m}n\nabla\Biggl(\frac{\triangle n}{n}-\frac{(\nabla n)^{2}}{2n^{2}}\Biggr)=-\textrm{g}n\nabla n$$
\begin{equation}\label{Twb dBEC bal imp eq with P2}+
\textrm{P}^{\beta}\nabla \textrm{E}^{\beta}_{ext}+\int \textrm{d}\textbf{r}'(\nabla \textrm{G}^{\beta\gamma}(|\textbf{r}-\textbf{r}'|)) \textrm{P}_{2}^{\beta\gamma}(\textbf{r},\textbf{r}',t),\end{equation}
where the particle concentration $n$ is defined in terms of many-particle wave function
\begin{equation}\label{Twb dBEC def density}n(\textbf{r},t)=\int \textrm{d}\textrm{R}_{N}\sum_{i}\delta(\textbf{r}-\textbf{r}_{i})\psi^{*}(\textrm{R},t)\psi(\textrm{R},t),\end{equation}
where $\textrm{d}\textrm{R}_{N}=\prod_{p=1}^{N}\textrm{d}\textbf{r}_{p}$ is an element of $3N$ volume.
The particle current $\textbf{j}=n\textbf{v}$ also has an explicit definition via the wave function (see for instance formula 4 in Ref. \cite{Andreev PRB 11}, formula 3 in Ref. \cite{Andreev RPJ 07}, and formula 10 in Ref. \cite{Andreev MPL 13}). The polarisation (density of electric dipole moment) $\textbf{P}(\textbf{r},t)$ arises in the second term on the right-hand side of equation (\ref{Twb dBEC bal imp eq with P2}). Definition of polarisation is
\begin{equation}\label{Twb dBEC polarisation Def} \textbf{P}(\textbf{r},t)=\int \textrm{d}\textrm{R}_{N}\sum_{i}\delta(\textbf{r}-\textbf{r}_{i})\textbf{d}_{i}\psi^{*}(\textrm{R},t)\psi(\textrm{R},t). \end{equation}
The second term on the right-hand side of equation (\ref{Twb dBEC bal imp eq with P2}) describes action of an inhomogeneous external electric field on dipolar BECs. The third term on the right-hand side presents the dipole-dipole interaction. It contains the Green function of interaction of electric dipoles
$$\textrm{G}^{\alpha\beta}(\textbf{r},\textbf{r}')=\partial^{\alpha}\partial^{\beta}\frac{1}{|\textbf{r}-\textbf{r}'|}$$
$$ =-\frac{\delta^{\alpha\beta}-3r^{\alpha}r^{\beta}/r^{2}}{r^{3}}-\frac{4\pi}{3}\delta^{\alpha\beta}\delta(\textbf{r})$$
\begin{equation}\label{Twb dBEC dd int Green funct} =-\frac{\delta^{\alpha\beta}-3r^{\alpha}r^{\beta}/r^{2}}{r^{3}}+\frac{1}{3}\delta^{\alpha\beta}\triangle\frac{1}{r}.\end{equation}
Equation (\ref{Twb dBEC bal imp eq with P2}) also contains the
two-particle polarisation
$$\textrm{P}_{2}^{\alpha\beta}(\textbf{r},\textbf{r}',t)=\int \textrm{d}\textrm{R}_{N}\sum_{i,j\neq i}\delta(\textbf{r}-\textbf{r}_{i})\delta(\textbf{r}'-\textbf{r}_{j})\times$$
\begin{equation}\label{Twb dBEC polarisation 2part Def}\times d_{i}^{\alpha}d_{j}^{\beta}\psi^{*}(\textrm{R},t)\psi(\textrm{R},t), \end{equation}
which is a second rank tensor. If we have system of align dipoles then polarisation (\ref{Twb dBEC polarisation Def}) simplifies to $\textbf{P}=d\textbf{l}n$, with $\textbf{l}$ direction of all dipoles, let us choose $\textbf{l}=\textbf{e}_{z}$. In this case formula (\ref{Twb dBEC polarisation 2part Def}) gives
\begin{equation}\label{Twb dBEC} \textrm{P}_{2}^{\alpha\beta}(\textbf{r},\textbf{r}',t)=d^{2}\delta^{\alpha z}\delta^{\beta z} n_{2}(\textbf{r},\textbf{r}',t), \end{equation}
where we have assumed that all dipoles parallel to $\textbf{e}_{z}$, and $n_{2}(\textbf{r},\textbf{r}',t)$ is the two-particle concentration
\begin{equation}\label{Twb dBEC concentration 2part Def} n_{2}(\textbf{r},\textbf{r}',t)=\int \textrm{d}\textrm{R}_{N}\sum_{i,j\neq i}\delta(\textbf{r}-\textbf{r}_{i})\delta(\textbf{r}'-\textbf{r}_{j})\psi^{*}(\textrm{R},t)\psi(\textrm{R},t). \end{equation}

Evolution of the many-particle wave function $\psi(\textrm{R},t)$ defined in 3N dimensional configurational space obeys the Schrodinger equation. Applying the quasi-static approximation to the electric dipoles we can write an explicit form of the Schrodinger equation
$$\imath\hbar\partial_{t}\psi(\textrm{R},t)=\biggl[\sum_{i}\Biggl(\frac{1}{2m_{i}}\hat{\textbf{p}}_{i}^{2}-\textbf{d}_{i}\textbf{E}_{i,ext}+V_{trap}(\textbf{r}_{i},t)\Biggr)$$
\begin{equation}\label{Twb dBEC Schrodinger}+\frac{1}{2}\sum_{i,j\neq i}\Biggl(U_{ij}-d_{i}^{\alpha}d_{j}^{\beta}G_{ij}^{\alpha\beta}\Biggr)\biggr]\psi(\textrm{R},t),\end{equation}
where $U_{ij}$ is the potential of the short-range interaction leading to the interaction constant $\textrm{g}$ (see the first term on the right-hand side of the Euler equation (\ref{Twb dBEC bal imp eq with P2})).

Let us consider properties of the two-particle concentration aimed to describe align dipoles. We will get back to general case described by the two-particle polarisation later.

The two-particle concentration was calculated in Refs. \cite{Andreev PRA08}, \cite{Andreev IJMP B 13} in terms of one-particle wave function, which describes the occupied quantum states. The two-particle concentration was found to be
$$n_2(\textbf{r},\textbf{r}',t)=n(\textbf{r},t)n(\textbf{r}',t)$$
\begin{equation}\label{Twb dBEC n2 expansion}
+|\rho(\textbf{r},\textbf{r}',t)|^{2}+\sum_{g}n_{g}(n_{g}-1)|\varphi_{g}(\textbf{r},t)|^{2}|\varphi_{g}(\textbf{r}',t)|^{2}
.\end{equation}

Here, the particle concentration
\begin{equation}\label{Twb dBEC n varphi}
n(\textbf{r},t)=\sum_{g}n_{g}\varphi_{g}^{*}(\textbf{r},t)\varphi_{g}(\textbf{r},t)
,\end{equation}
and the density matrix
\begin{equation}\label{Twb dBEC rho varphi}\rho(\textbf{r},\textbf{r}',t)=\sum_{g}n_{g}\varphi_{g}^{*}(\textbf{r},t)\varphi_{g}(\textbf{r}',t),\end{equation}
in terms of the arbitrary single-particle wave functions $\varphi_{g}(\textbf{r},t)$.

The first two terms in formula (\ref{Twb dBEC n2 expansion}) represents the
particles situating in two different quantum states, while the
third term is referred to particles in the same quantum state.
Therefore, for the particles in the BEC state, it is sufficient to
take into account the third term in formula
(\ref{Twb dBEC n2 expansion}). In consideration of the system of bosons
with the temperature differing from zero, where the certain number
of the particles is out of the condensate, the first two summands of
formula (\ref{Twb dBEC n2 expansion}) gives the contribution both in the
case of interaction of excited particles with each other and in
the case of their interaction with the particles appearing in the
BEC state. In this case, the third term of formula
(\ref{Twb dBEC n2 expansion}) gives the contribution in the interaction
both between the particles appearing in the BEC state and between
the excited particles appearing in the same quantum state.

The full concentration can be separated on two parts $n=n_{B}+n_{n}$, where we have used the notions $n_{B}(\textbf{r},t)$ for the
concentration of particles situating in the BEC state and
$n_{n}(\textbf{r},t)$ for the concentration of excited particles. Hence the product of concentration in formula (\ref{Twb dBEC n2 expansion}), formally, appears to be
$$n(\textbf{r},t)n(\textbf{r}',t)=\biggl[n_{B}(\textbf{r},t)n_{B}(\textbf{r}',t)+n_{B}(\textbf{r},t)n_{n}(\textbf{r}',t)$$
\begin{equation}\label{Twb dBEC product of n n formal} +n_{n}(\textbf{r},t)n_{B}(\textbf{r}',t)+n_{n}(\textbf{r},t)n_{n}(\textbf{r}',t)\biggr]_{formal},\end{equation}
but we cannot have product of functions describing particles in the same quantum state. Hence we should exclude the first term on the right-hand side of formula (\ref{Twb dBEC product of n n formal}). Finally we obtain
$$n(\textbf{r},t)n(\textbf{r}',t)= n_{B}(\textbf{r},t)n_{n}(\textbf{r}',t)$$
\begin{equation}\label{Twb dBEC product of n n formal} +n_{n}(\textbf{r},t)n_{B}(\textbf{r}',t)+n_{n}(\textbf{r},t)n_{n}(\textbf{r}',t).\end{equation}

Similar picture we have for the second term in formula (\ref{Twb dBEC n2 expansion})
$$|\rho(\textbf{r},\textbf{r}',t)|^{2} =\rho(\textbf{r},\textbf{r}',t)\rho(\textbf{r}',\textbf{r},t)$$
\begin{equation}\label{Twb dBEC} =\rho_{B}^{*}\rho_{n}+\rho_{n}^{*}\rho_{B}+\rho_{n}^{*}\rho_{n},\end{equation}
where we have used $\rho(\textbf{r}',\textbf{r},t)=\rho^{*}(\textbf{r},\textbf{r}',t)$.

The first term in formula (\ref{Twb dBEC n2 expansion}) corresponds to the self-consistent field approximation. Second and third terms of
formula (\ref{Twb dBEC n2 expansion}) represent the part of the force field
caused by the exchange interaction. Substituting expression
(\ref{Twb dBEC n2 expansion}) into the force field of dipole-dipole interaction of the boson system, we derive the formula
$$\textbf{F}(\textbf{r},t)\equiv \int d\textbf{r}'(\nabla \textrm{G}^{\beta\gamma}(|\textbf{r}-\textbf{r}'|)) \textrm{P}_{2}^{\beta\gamma}(\textbf{r},\textbf{r}',t)$$
$$=d^{2}\int d\textbf{r}'(\nabla \textrm{G}^{zz}(|\textbf{r}-\textbf{r}'|))\biggl(n_{B}(\textbf{r},t)n_{n}(\textbf{r}',t)$$
$$+n_{n}(\textbf{r},t)n_{B}(\textbf{r}',t)+n_{n}(\textbf{r},t)n_{n}(\textbf{r}',t)$$
$$+\rho_{B}^{*}(\textbf{r},\textbf{r}',t)\rho_{n}(\textbf{r},\textbf{r}',t) +\rho_{n}^{*}(\textbf{r},\textbf{r}',t)\rho_{B}(\textbf{r},\textbf{r}',t)$$
\begin{equation}\label{Twb dBEC ForceField boz and sm exitation} +\rho_{n}^{*}(\textbf{r},\textbf{r}',t)\rho_{n}(\textbf{r},\textbf{r}',t) +\sum_{g}n_{g}(n_{g}-1)|\varphi_{g}(\textbf{r},t)|^{2}|\varphi_{g}(\textbf{r}',t)|^{2}\biggr),\end{equation}
where $G^{zz}(\xi)=\textrm{G}^{\alpha\beta}(\xi)\delta^{z\alpha}\delta^{z\beta}=(-\frac{\delta^{\alpha\beta}}{\xi^{3}} +\frac{3\xi^{\alpha}\xi^{\beta}}{\xi^{5}} -\frac{4\pi}{3}\delta^{\alpha\beta}\delta(\xi))\delta^{z\alpha}\delta^{z\beta}$ $=-\frac{1}{\xi^{3}}+\frac{3(\xi^{z})^{2}}{\xi^{5}} -\frac{4\pi}{3}\delta(\xi)$ is an explicit form of the zz matrix element of the Green function of the electric dipole interaction.

Dropping contribution of the excited states we obtain the force field of dipole-dipole interaction for boson systems in the BEC state
$$\textbf{F}(\textbf{r},t)=d^{2}\int d\textbf{r}'(\nabla \textrm{G}^{zz}(|\textbf{r}-\textbf{r}'|))\times$$
\begin{equation}\label{Twb dBEC ForceField dd BEC} \times\sum_{g=g_{0}}n_{g}(n_{g}-1)|\varphi_{g}(\textbf{r},t)|^{2}|\varphi_{g}(\textbf{r}',t)|^{2},\end{equation}
where $g_{0}$ is an index of the state with lower energy.

Formula (\ref{Twb dBEC ForceField dd BEC}) appears from the last term of formula (\ref{Twb dBEC ForceField boz and sm exitation}). To be more precise we should note that formula (\ref{Twb dBEC ForceField dd BEC}) corresponds to a single term in the sum presented by the last term of formula (\ref{Twb dBEC ForceField boz and sm exitation}).

Further manipulations with the force of electric dipole interaction of dipolar particles in the BEC state (\ref{Twb dBEC ForceField dd BEC}) give
$$\textbf{F}(\textbf{r},t)= d^{2}\int d\textbf{r}'(\nabla \textrm{G}^{zz}(|\textbf{r}-\textbf{r}'|))\times$$
$$\times n_{g_{0}}(n_{g_{0}}-1)|\varphi_{g_{0}}(\textbf{r},t)|^{2}|\varphi_{g_{0}}(\textbf{r}',t)|^{2}$$
$$\approx d^{2}\int d\textbf{r}'(\nabla \textrm{G}^{zz}(|\textbf{r}-\textbf{r}'|))(n_{g_{0}}|\varphi_{g_{0}}(\textbf{r},t)|^{2})(n_{g_{0}}|\varphi_{g_{0}}(\textbf{r}',t)|^{2})$$
\begin{equation}\label{Twb dBEC ForceField dd BEC calculation} \approx d^{2}\int d\textbf{r}'(\nabla \textrm{G}^{zz}(|\textbf{r}-\textbf{r}'|))n(\textbf{r},t) n(\textbf{r}',t).\end{equation}
Calculations in formula (\ref{Twb dBEC ForceField dd BEC calculation}) are in accordance with formula (\ref{Twb dBEC n varphi}), which gives $n_{B}=n_{g_{0}}|\varphi_{g_{0}}(\textbf{r},t)|^{2}$ in the case of all particles located in the BEC state.

Finally we have
\begin{equation}\label{Twb dBEC ForceField correl final form} \textbf{F}(\textbf{r},t)= d^{2}n(\textbf{r},t) \int d\textbf{r}'(\nabla \textrm{G}^{zz}(|\textbf{r}-\textbf{r}'|)) n(\textbf{r}',t).\end{equation}
This result looks like outcome of the formal application of the self-consistent field approximation, but it has different meaning. Behavior of the exchange correlations in ultracold fermions is rather different, analysis of the Coulomb exchange interaction in quantum plasmas of degenerate electrons was presented in Ref. \cite{Andreev 1403 exchange}.

The self-consistent field approximation can not be applied to the long-range interacting particles in the BEC state. However we can use the similarity of the force field (\ref{Twb dBEC ForceField correl final form}) with the formal application of the self-consistent field approximation. This similarity allows to introduce the internal electric field created by the dipoles of the medium, so the force field (\ref{Twb dBEC ForceField correl final form}) reappear in the following form
\begin{equation}\label{Twb dBEC ForceField short form} \textbf{F}(\textbf{r},t)= d n(\textbf{r},t) \nabla(\textbf{l}\cdot\textbf{E}),\end{equation}
where
\begin{equation}\label{Twb dBEC el field zz} \textbf{E}=d\nabla\int \textrm{d}\textbf{r}'(\textrm{G}^{zz}(|\textbf{r}-\textbf{r}'|)) n(\textbf{r}',t) \end{equation}

The internal electric field (\ref{Twb dBEC el field zz}) satisfies the Maxwell equations
\begin{equation}\label{Twb dBEC field good div}\nabla\cdot\textbf{E}(\textbf{r},t)=-4\pi \nabla\cdot\textbf{P}(\textbf{r},t)=-4\pi (\textbf{d}\cdot\nabla) n(\textbf{r},t),\end{equation}
and
\begin{equation}\label{Twb dBEC field good curl}\nabla\times\textbf{E}(\textbf{r},t)=0.\end{equation}

The Euler equation for dipolar BECs (\ref{Twb dBEC bal imp eq with P2}) now can be presented as follows
$$mn(\partial_{t}+\textbf{v}\cdot\nabla)\textbf{v}-\frac{\hbar^{2}}{4m}n\nabla\Biggl(\frac{\triangle n}{n}-\frac{(\nabla n)^{2}}{2n^{2}}\Biggr)$$
\begin{equation}\label{Twb dBEC bal imp eq non Int} =-\textrm{g}n\nabla n+
d n \nabla (\textbf{l}\cdot\textbf{E}).\end{equation}

Equations (\ref{Twb dBEC cont eq from GP}), (\ref{Twb dBEC field good div}), (\ref{Twb dBEC field good curl}), and (\ref{Twb dBEC bal imp eq non Int}) form a closed set of equations, which can be applied to the analysis of dipolar BECs.

Formulae (\ref{Twb dBEC ForceField correl final form}) and (\ref{Twb dBEC bal imp eq non Int}) are among the main results of this paper. We should stress attention readers that they are obtained at quantum exchange interaction. These formulae do not related to the self-consistent field approximation, but their form coincides with the results of application of the self-consistent field approximation. This twisted behavior of the model of dipolar BECs allows to justify equations applied in Refs. \cite{Andreev MPL 13} and \cite{Andreev EPJ D 14}. It also gives partial justification of papers \cite{Andreev EPJ D 13}, \cite{Andreev TransDipBEC12}, and \cite{Andreev RPJ 12}, where the evolution of the dipole directions is also  considered. The full justification of these papers will be given in the next section of this paper, but general idea is the same as we have in this section.

Since all particles are located in the same quantum state, so they are described by the sam,e single particle wave function. Consequently their velocity field is potential $\textbf{v}(\textbf{r},t)=\nabla\phi(\textbf{r},t)$. Hence we can derive corresponding non-linear Schrodinger equation for the macroscopic wave function (the order parameter, or the wave function in the medium) from the continuity equation (\ref{Twb dBEC cont eq from GP}) the Euler equation (\ref{Twb dBEC bal imp eq non Int}).

The macroscopic wave function is defined in terms of hydrodynamic variables
\begin{equation}\label{Twb dBEC WF def} \Phi(\textbf{r},t)=\sqrt{n(\textbf{r},t)}\exp(\imath m\phi(\textbf{r},t)/\hbar),\end{equation}
where $\phi(\textbf{r},t)$ is the potential of the velocity field $\textbf{v}(\textbf{r},t)$.

Differentiating function (\ref{Twb dBEC WF def}) with respect to time we find the non-linear Schrodinger equation
\begin{equation}\label{Twb dBEC nlse int polariz some frame non Int}\imath\hbar\partial_{t}\Phi(\textbf{r},t)=\Biggl(-\frac{\hbar^{2}}{2m}\triangle +\textrm{g}\mid\Phi(\textbf{r},t)\mid^{2}-\textbf{d}\cdot\textbf{E}\Biggr)\Phi(\textbf{r},t),\end{equation}
which is the Gross-Pitaevskii equation for dipolar BECs. Equation (\ref{Twb dBEC nlse int polariz some frame non Int}) is an equivalent of the set of quantum hydrodynamic equations (\ref{Twb dBEC cont eq from GP}) and (\ref{Twb dBEC bal imp eq non Int}). As we can see equation (\ref{Twb dBEC nlse int polariz some frame non Int}) is a non-integral Gross-Pitaevskii equation.

Similar result appears for the force field of Coulomb interaction between Cooper pairs of electrons in superconductors:
\begin{equation}\label{Twb dBEC ForceField Cooper pairs} \textbf{F}(\textbf{r},t)= (2e)^{2}n(\textbf{r},t) \int \textrm{d}\textbf{r}'(\nabla \textrm{G}(|\textbf{r}-\textbf{r}'|)) n(\textbf{r}',t),\end{equation}
where $\textrm{G}=1/|\textbf{r}-\textbf{r}'|$ is the Green function of the Coulomb interaction. Formula arises as the result of calculation of the two-particle concentration (\ref{Twb dBEC concentration 2part Def}) and (\ref{Twb dBEC n2 expansion}) with no references to the self-consistent field approximation. The force field (\ref{Twb dBEC ForceField Cooper pairs}) corresponds to the Landau-Ginzburg equation \cite{Rosenstein RMP 10}.

We have illustrated the strange behavior of the dipolar BECs on an example of the electric dipolar BECs, but we have similar picture for the magnetic dipolar BECs. Let us also mention that difference in behavior of the electric and magnetic dipolar BECs was described in Refs. \cite{Andreev EPJ D 14} and \cite{Andreev EPJ D 13}.

\subsection{Two dimensional dipolar BECs of aligned dipoles}

Two dimensional dipolar BECs were recently considered in Refs. \cite{Lu_Shlyapnikov arX_14} and \cite{Boudjemaa PRA 13}. The standard model of dipolar BECs does not employ the delta function term in the potential of electric dipole interaction \cite{Santos PRL 00}, \cite{Goral PRA 00}, \cite{Yi PRA 00}. However it creates some problem at transition to two-dimensional samples.

Since the two dimensional plane-like structure of dipoles is located in three dimensional space. Hence dipoles do not have to be parallel to the plane, but they can be directed at an angle to the plane, or they can be directed perpendicular to the plane. Consequently the potential of dipole-dipole interaction (\ref{Twb dBEC dd int Green funct}) in two-dimensional sample can be written as $$U_{dd}(2D)=-d^{\alpha}d^{\beta}[\partial^{\alpha}\partial^{\beta}\frac{1}{|\textbf{r}-\textbf{r}'|}]|_{z=z'=0}$$ $$=-[d^{\alpha}d^{\beta}]|_{inplane}[\partial^{\alpha}\partial^{\beta}\frac{1}{|\textbf{r}-\textbf{r}'|}]|_{z=z'=0}$$ $$+d_{z}^{2}(-\frac{1}{|\textbf{r}-\textbf{r}'|^{3}}+\frac{1}{3}\triangle\frac{1}{r})|_{z=z'=0}$$.

Let us consider the last term containing the projection of the dipoles on z-axis, which is perpendicular to the plane. As a possible way to deal with two-dimensional samples we can rewrite this term as follows
\begin{equation}\label{Twb dBEC potential dz dz} U_{d_{z}d_{z}}(2D)=d_{z}^{2}\biggl(-\frac{1}{|\textbf{r}_{2D}-\textbf{r}'_{2D}|^{3}}+\frac{1}{3}\triangle\frac{1}{r_{2D}}\biggr), \end{equation}
where we have substituted $z=z'=0$ before differentiating $\frac{1}{r}$. We can notice that in the two-dimensional case $\triangle\frac{1}{r_{2D}}=\frac{1}{r_{2D}^{3}}$, where we included $\nabla \textbf{r}_{2D}=2$. Consequently we can present a short form of potential of dipole-dipole interaction of dipoles perpendicular to the plane \ref{Twb dBEC potential dz dz} in the next form
\begin{equation}\label{Twb dBEC} U_{d_{z}d_{z}}(2D)=-\frac{2}{3}d_{z}^{2}\frac{1}{|\textbf{r}_{2D}|},\end{equation}
which differs by multiplier $2/3$ from the similar result obtain with no application of the delta function term in the original potential of the electric dipole interaction (\ref{Twb dBEC dd int Green funct}).

The Fourier transform of the inplane part of the electric dipole potential reads $U_{inplane}(\textbf{k})=(\textbf{d}\textbf{k})^{2}\frac{2\pi}{k}$. The Fourier transform of the part of the electric dipole potential related to dipoles perpendicular to the plane is $U_{d_{z}d_{z}}(\textbf{k})=4\pi k/3$.

\section{Dipolar BEC with dipole direction evolution}

If we do not consider approximation of the aligned dipoles and include the evolution of the dipole directions we need to calculate the two-particle polarisation $\textrm{P}_{2}^{\alpha\beta}(\textbf{r},\textbf{r}',t)$ instead of the two-particle concentration $n_{2}(\textbf{r},\textbf{r}',t)$.

Our calculation of the two-particle polarisation $\textrm{P}_{2}^{\alpha\beta}(\textbf{r},\textbf{r}',t)$ gives
$$ \textrm{P}_2^{\alpha\beta}(\textbf{r},\textbf{r}',t)=\textrm{P}^{\alpha}(\textbf{r},t)\textrm{P}^{\beta}(\textbf{r}',t)$$
$$+\frac{1}{2}\biggl(\Gamma^{\alpha}(\textbf{r},\textbf{r}',t)(\Gamma^{\beta}(\textbf{r},\textbf{r}',t))^{*}+c.c.\biggr)$$
\begin{equation}\label{Twb dBEC P2 expansion}
+\sum_{g}n_{g}(n_{g}-1)d_{g}^{\alpha}d_{g}^{\beta}|\varphi_{g}(\textbf{r},t)|^{2}|\varphi_{g}(\textbf{r}',t)|^{2}
,\end{equation}
where
\begin{equation}\label{Twb dBEC P varphi}
\textbf{P}(\textbf{r},t)=\sum_{g}n_{g}\textbf{d}_{g}\varphi_{g}^{*}(\textbf{r},t)\varphi_{g}(\textbf{r},t)
,\end{equation}
and
\begin{equation}\label{Twb dBEC Gamma varphi} \Gamma^{\alpha}(\textbf{r},\textbf{r}',t)=\sum_{g}n_{g}d_{g}^{\alpha}\varphi_{g}^{*}(\textbf{r},t)\varphi_{g}(\textbf{r}',t),\end{equation}
with application of the
arbitrary single-particle wave functions $\varphi_{g}(\textbf{r},t)$ presented above. Dipole moments have subindex $g$, because they different in different quantum states.

Dropping contribution of the excited states we obtain the force field of dipole-dipole interaction for boson systems in the BEC state
$$\textbf{F}(\textbf{r},t)=\int \textrm{d}\textbf{r}'(\nabla \textrm{G}^{\beta\gamma}(|\textbf{r}-\textbf{r}'|))\times$$
\begin{equation}\label{Twb dBEC ForceField dd BEC with P evol} \times\sum_{g=g_{0}}n_{g}(n_{g}-1)d_{g}^{\beta}d_{g}^{\gamma}|\varphi_{g}(\textbf{r},t)|^{2}|\varphi_{g}(\textbf{r}',t)|^{2},\end{equation}
where $g_{0}$ is an index of the state with lower energy.

Further manipulations give
$$\textbf{F}(\textbf{r},t)= \int \textrm{d}\textbf{r}'(\nabla \textrm{G}^{\beta\gamma}(|\textbf{r}-\textbf{r}'|))\times$$
$$\times n_{g_{0}}(n_{g_{0}}-1))d_{g}^{\beta}d_{g}^{\gamma}|\varphi_{g_{0}}(\textbf{r},t)|^{2}|\varphi_{g_{0}}(\textbf{r}',t)|^{2}$$
$$\approx \int \textrm{d}\textbf{r}'(\nabla \textrm{G}^{\beta\gamma}(|\textbf{r}-\textbf{r}'|)))\times$$
$$\times d_{g}^{\beta}d_{g}^{\gamma}(n_{g_{0}}|\varphi_{g_{0}}(\textbf{r},t)|^{2})(n_{g_{0}}|\varphi_{g_{0}}(\textbf{r}',t)|^{2})$$
\begin{equation}\label{Twb dBEC ForceField dd BEC calculation P evol} \approx \int \textrm{d}\textbf{r}'(\nabla \textrm{G}^{\beta\gamma}(|\textbf{r}-\textbf{r}'|))\textrm{P}_{B}^{\beta}(\textbf{r},t) \textrm{P}_{B}^{\gamma}(\textbf{r}',t),\end{equation}
in accordance with formula (\ref{Twb dBEC P varphi}), which gives $\textbf{P}_{B}=n_{g_{0}}\textbf{d}_{g_{0}}|\varphi_{g_{0}}(\textbf{r},t)|^{2}$ in the case of all particles located in the BEC state.

This calculation leads to the following force field
\begin{equation}\label{Twb dBEC ForceField correl final form P evol} \textbf{F}(\textbf{r},t)= \textrm{P}^{\beta}(\textbf{r},t) \int \textrm{d}\textbf{r}'(\nabla \textrm{G}^{\beta\gamma}(|\textbf{r}-\textbf{r}'|)) \textrm{P}^{\gamma}(\textbf{r}',t).\end{equation}
This result coincides with the formal application of the self-consistent field approximation in the Euler equation for dipolar BECs with the dipolar direction evolution, but it is based on the dipole-dipole exchange interaction.

Similarly to the aligned dipoles (see formulae (\ref{Twb dBEC ForceField short form})-(\ref{Twb dBEC bal imp eq non Int})) we can use similarity of our result (\ref{Twb dBEC ForceField correl final form P evol}) with the self-consistent field approximation and introduce the electric field created by dipoles
\begin{equation}\label{Twb dBEC El field Int with P} \textrm{E}^{\alpha}(\textbf{r},t)= \int \textrm{d}\textbf{r}'(\textrm{G}^{\alpha\beta}(|\textbf{r}-\textbf{r}'|)) \textrm{P}^{\beta}(\textbf{r}',t). \end{equation}
In this case the force field (\ref{Twb dBEC ForceField correl final form P evol}) can be rewritten as $\textbf{F}=\textrm{P}^{\beta}\nabla \textrm{E}^{\beta}$. So we can rewrite the Euler equation as follows
$$mn(\partial_{t}+\textbf{v}\cdot\nabla)\textbf{v}-\frac{\hbar^{2}}{4m}n\nabla\Biggl(\frac{\triangle n}{n}-\frac{(\nabla n)^{2}}{2n^{2}}\Biggr)$$
\begin{equation}\label{Twb dBEC bal imp eq non Int with P dir evol} =-\textrm{g}n\nabla n+
\textrm{P}^{\beta}\nabla \textrm{E}^{\beta}.\end{equation}

The electric field satisfies the Maxwell equation
\begin{equation}\label{Twb dBEC field good div with P dir evol}\nabla\cdot\textbf{E}(\textbf{r},t)=-4\pi \nabla\cdot\textbf{P}(\textbf{r},t),\end{equation}
and
\begin{equation}\label{Twb dBEC field good curl with P dir evol}\nabla\times\textbf{E}(\textbf{r},t)=0.\end{equation}

\subsection{Polarisation evolution}

In this section we consider the dipole direction evolution. Consequently we can not give simple representation of the polarisation $\textbf{P}$ in terms of the particle concentration $n$. Hence the set of continuity (\ref{Twb dBEC cont eq from GP}), Euler (\ref{Twb dBEC bal imp eq non Int with P dir evol}), and Maxwell (\ref{Twb dBEC field good div with P dir evol}), (\ref{Twb dBEC field good curl with P dir evol}) equations is not a closed set. Now we need to find equation for polarisation evolution. This equation can be derived by differentiating the definition of polarisation (\ref{Twb dBEC polarisation Def}) with respect to time, analogously to derivation of the continuity equation by the differentiating of the particle concentration (\ref{Twb dBEC def density}) with respect to time. Let us mention that the time derivative of the many-particle wave function can be taken from the Schrodinger equation (\ref{Twb dBEC Schrodinger}). After some straightforward calculations one can find
the equation of polarization evolution
\begin{equation}\label{Twb dBEC eq polarization} \partial_{t}\textrm{P}^{\alpha}(\textbf{r},t)+\partial^{\beta}\textrm{R}^{\alpha\beta}(\textbf{r},t)=0,\end{equation}
where $\textrm{R}^{\alpha\beta}(\textbf{r},t)$ is the current of polarization \cite{Andreev EPJ D 13}, \cite{Andreev TransDipBEC12}, \cite{Andreev RPJ 13}, \cite{Andreev PRB 11}, \cite{Andreev RPJ 12}. Its explicit  form is
$$\textrm{R}^{\alpha\beta}(\textbf{r},t)=\int \textrm{d}\textrm{R}\sum_{i}\delta(\textbf{r}-\textbf{r}_{i})\frac{d_{i}^{\alpha}}{2m_{i}}\times$$
\begin{equation}\label{Twb dBEC def of current of polarization} \times\biggl(\psi^{*}(\textrm{R},t)(\textrm{D}_{i}^{\beta}\psi(\textrm{R},t)) +(\textrm{D}_{i}^{\beta}\psi(\textrm{R},t))^{*}\psi(\textrm{R},t)\biggr).\end{equation}

The equation of polarisation evolution (\ref{Twb dBEC eq polarization}) does not contain any information about the influence of the interaction on the polarisation evolution. So, following Refs. \cite{Andreev EPJ D 13}, \cite{Andreev TransDipBEC12}, \cite{Andreev RPJ 13}, \cite{Andreev PRB 11} we derive the polarisation current $\textrm{R}^{\alpha\beta}(\textbf{r},t)$ evolution equation
$$\partial_{t}\textrm{R}^{\alpha\beta}+\partial^{\gamma}\biggl(\textrm{R}^{\alpha\beta}v^{\gamma} +\textrm{R}^{\alpha\gamma}v^{\beta}-\textrm{P}^{\alpha}v^{\beta}v^{\gamma}\biggr)$$
$$-\frac{\hbar^{2}}{4m^{2}}\partial_{\beta}\triangle \textrm{P}^{\alpha}
+\frac{\hbar^{2}}{8m^{2}}\partial^{\gamma}\biggl(\frac{\partial_{\beta}\textrm{P}^{\alpha}\cdot\partial_{\gamma}n}{n} +\frac{\partial_{\gamma}\textrm{P}^{\alpha}\cdot\partial_{\beta}n}{n}\biggr)
$$
$$=-\frac{1}{2m}\textrm{g}\partial^{\beta}\textrm{P}^{\alpha}(\textbf{r},\textbf{r},t) $$
\begin{equation}\label{Twb dBEC eq for pol current gen with two part} +\frac{1}{m}\int \textrm{d}\textbf{r}'(\partial^{\beta} \textrm{G}^{\gamma\delta}(|\textbf{r}-\textbf{r}'|)) \textrm{D}_{2}^{\alpha\gamma\delta}(\textbf{r},\textbf{r}',t), \end{equation}
where we have included the explicit form of representation of the third-rank tensor of the flux of polarisation current. This tensor splits on three parts. The first of them is presented by the second group of terms on the left-hand side of equation (\ref{Twb dBEC eq for pol current gen with two part}). It is related to the motion of the local centre of mass and contain the velocity field $\textbf{v}$. It is an analog of $(\textbf{v}\cdot\nabla) \textbf{v}$ in the Euler equation. The second part of the flux of polarisation current is the quantum part presented by the third and fourth terms in the polarisation current evolution equation. These terms are proportional to the square of the Planck constant. The third part is related to the thermal motion. In the zero temperature limit, corresponding to the BEC dynamics, this term equals to zero, so it is not presented in equation (\ref{Twb dBEC eq for pol current gen with two part}).

Equation (\ref{Twb dBEC eq for pol current gen with two part}) contains two two-particle hydrodynamic functions, here we present their definitions arising during derivation of equation (\ref{Twb dBEC eq for pol current gen with two part})
\begin{equation}\label{Twb dBEC definition of trace nP_2} \textrm{P}^{\alpha}(\textbf{r},\textbf{r},t)=Tr(\textrm{P}^{\alpha}(\textbf{r},\textbf{r}',t)),\end{equation}
with
$$\textrm{P}^{\alpha}(\textbf{r},\textbf{r}',t)=\int \textrm{d}\textrm{R}_{N}\sum_{i,j\neq i}\delta(\textbf{r}-\textbf{r}_{i})\delta(\textbf{r}'-\textbf{r}_{j})\times$$
\begin{equation}\label{Twb dBEC definition of trace nP_2}\times d_{i}^{\alpha}\psi^{*}(\textrm{R},t)\psi(\textrm{R},t), \end{equation}
and
$$\textrm{D}_{2}^{\alpha\beta\gamma}(\textbf{r},\textbf{r}',t)=\int \textrm{d}\textrm{R}_{N}\sum_{i,j\neq i}\delta(\textbf{r}-\textbf{r}_{i})\delta(\textbf{r}'-\textbf{r}_{j})\times$$
\begin{equation}\label{Twb dBEC D--polarisation 2part Def}\times d_{i}^{\alpha}d_{i}^{\beta}d_{j}^{\gamma}\psi^{*}(\textrm{R},t)\psi(\textrm{R},t). \end{equation}

Function $\textrm{D}_{2}^{\alpha\beta\gamma}(\textbf{r},\textbf{r}',t)$ can be considered similarly to the two-particle concentration $n_{2}(\textbf{r},\textbf{r}',t)$ (see formulae (\ref{Twb dBEC concentration 2part Def}) and (\ref{Twb dBEC n2 expansion})), and the two-particle polarisation $\textrm{P}_{2}^{\alpha\beta}(\textbf{r},\textbf{r}',t)$ (see formulae (\ref{Twb dBEC polarisation 2part Def}) and (\ref{Twb dBEC P2 expansion})). Hence we can find that for particles collected in the BEC state function $\textrm{D}_{2}^{\alpha\beta\gamma}(\textbf{r},\textbf{r}',t)$ arises in the following form
$$ \textrm{D}_{2}^{\alpha\beta\gamma}(\textbf{r},\textbf{r}',t)$$
\begin{equation}\label{Twb dBEC Dabc2 expansion}
=\sum_{g=g_{0}}n_{g}(n_{g}-1)d_{g}^{\alpha}d_{g}^{\beta}d_{g}^{\gamma}|\varphi_{g}(\textbf{r},t)|^{2}|\varphi_{g}(\textbf{r}',t)|^{2}
.\end{equation}

To interpret this formula we need to consider the first term of the expansion of function $\textrm{D}_{2}^{\alpha\beta\gamma}(\textbf{r},\textbf{r}',t)$ corresponding to the self-consistent field approximation,  which is $\textrm{D}_{2}^{\alpha\beta\gamma}(\textbf{r},\textbf{r}',t)\rightarrow \textrm{D}^{\alpha\beta}(\textbf{r},t)\cdot \textrm{P}^{\gamma}(\textbf{r}',t)$.
Definition of function $D^{\alpha\beta}$ occurs here is
\begin{equation}\label{Twb dBEC definition of D}\textrm{D}^{\alpha\beta}(\textbf{r},t)=\int \textrm{d}\textrm{R}\sum_{i}\delta(\textbf{r}-\textbf{r}_{i})d_{i}^{\alpha}d_{i}^{\beta}\psi^{*}(\textrm{R},t)\psi(\textrm{R},t).\end{equation}
An approximate formula for this function has been used in literature \cite{Andreev EPJ D 13}, \cite{Andreev RPJ 13}, \cite{Andreev PRB 11}, \cite{Andreev RPJ 12}
\begin{equation}\label{Twb dBEC appr for D} \textrm{D}^{\alpha\beta}(\textbf{r},t)=\sigma\frac{\textrm{P}^{\alpha}(\textbf{r},t)\textrm{P}^{\beta}(\textbf{r},t)}{n(\textbf{r},t)}.\end{equation}

We can analyze formula (\ref{Twb dBEC appr for D}) for dipolar BECs.
In terms of functions $\varphi_{g}(\textbf{r},t)$ function $\textrm{D}^{\alpha\beta}$ can be written as
\begin{equation}\label{Twb dBEC D varphi}
\textrm{D}^{\alpha\beta}(\textbf{r},t)=\sum_{g}n_{g}d_{g}^{\alpha}d_{g}^{\beta}\varphi_{g}^{*}(\textbf{r},t)\varphi_{g}(\textbf{r},t)
\end{equation}
(for polarisation $\textrm{P}^{\gamma}(\textbf{r}',t)$ see formula (\ref{Twb dBEC P varphi})).

Function $\textrm{D}^{\alpha\beta}$ in the limit of BEC is
\begin{equation}\label{Twb dBEC D varphi BEC}
\textrm{D}^{\alpha\beta}(\textbf{r},t)=n_{g_{0}}d_{g_{0}}^{\alpha}d_{g_{0}}^{\beta}\varphi_{g_{0}}^{*}(\textbf{r},t)\varphi_{g_{0}}(\textbf{r},t).
\end{equation}
Comparing formulae (\ref{Twb dBEC appr for D}) and (\ref{Twb dBEC D varphi BEC}) we find that numerical coefficient in formula (\ref{Twb dBEC appr for D}) should be equal to one in the limit of the zeroth temperature bosons $\sigma_{BEC}=1$.

Similarly we can write function $\textrm{D}_{2}^{\alpha\beta\gamma}(\textbf{r},\textbf{r}',t)$ in term of function $\textrm{D}^{\alpha\beta}$, or in terms of polarisation only, at our choice
$$\textrm{D}_{2}^{\alpha\beta\gamma}(\textbf{r},\textbf{r}',t)=\textrm{D}^{\alpha\beta}(\textbf{r},t) \textrm{P}^{\gamma}(\textbf{r}',t)$$
\begin{equation}\label{Twb dBEC Dabc2 expansion final} =\frac{1}{n(\textbf{r},t)}\textrm{P}^{\alpha}(\textbf{r},t)\textrm{P}^{\beta}(\textbf{r},t)\textrm{P}^{\gamma}(\textbf{r}',t).\end{equation}

We have finished discussion of function $\textrm{D}_{2}^{\alpha\beta\gamma}(\textbf{r},\textbf{r}',t)$, so we can substitute of final result (\ref{Twb dBEC Dabc2 expansion final}) in the equation of the polarisation current evolution (\ref{Twb dBEC eq for pol current gen with two part})
$$\partial_{t}\textrm{R}^{\alpha\beta}+\partial^{\gamma}\biggl(\textrm{R}^{\alpha\beta}v^{\gamma}+\textrm{R}^{\alpha\gamma}v^{\beta}-\textrm{P}^{\alpha}v^{\beta}v^{\gamma}\biggr)$$
$$-\frac{\hbar^{2}}{4m^{2}}\partial_{\beta}\triangle \textrm{P}^{\alpha}
+\frac{\hbar^{2}}{8m^{2}}\partial^{\gamma}\biggl(\frac{\partial_{\beta}\textrm{P}^{\alpha}\cdot\partial_{\gamma}n}{n}+\frac{\partial_{\gamma}\textrm{P}^{\alpha}\cdot\partial_{\beta}n}{n}\biggr)
$$
\begin{equation}\label{Twb dBEC eq for pol current gen single part Integral} =-\frac{1}{2m}\textrm{g}\partial^{\beta}(n\textrm{P}^{\alpha})+\frac{1}{m}\textrm{D}^{\alpha\gamma}\partial^{\beta}\int \textrm{d}\textbf{r}'(\textrm{G}^{\gamma\delta}(|\textbf{r}-\textbf{r}'|)) \textrm{P}^{\delta}(\textbf{r}',t), \end{equation}
with function $\textrm{D}^{\alpha\beta}=\textrm{P}^{\alpha}\textrm{P}^{\beta}/n$.

Non-integral form of the polarisation current evolution equation can be found with traditional introduction of the electric field created dy dipoles
$$\partial_{t}\textrm{R}^{\alpha\beta}+\partial^{\gamma}\biggl(\textrm{R}^{\alpha\beta}v^{\gamma}+\textrm{R}^{\alpha\gamma}v^{\beta}-\textrm{P}^{\alpha}v^{\beta}v^{\gamma}\biggr)$$
$$-\frac{\hbar^{2}}{4m^{2}}\partial_{\beta}\triangle \textrm{P}^{\alpha}
+\frac{\hbar^{2}}{8m^{2}}\partial^{\gamma}\biggl(\frac{\partial_{\beta}\textrm{P}^{\alpha}\cdot\partial_{\gamma}n}{n} +\frac{\partial_{\gamma}\textrm{P}^{\alpha}\cdot\partial_{\beta}n}{n}\biggr)
$$
\begin{equation}\label{Twb dBEC eq for pol current gen single part with NO Integral} =-\frac{1}{2m}\textrm{g}\partial^{\beta}(n\textrm{P}^{\alpha}) +\frac{1}{m}\frac{\textrm{P}^{\alpha}\textrm{P}^{\gamma}}{n}\partial^{\beta}\textrm{E}^{\gamma}, \end{equation}
where the electric field $\textbf{E}$ obeys the Maxwell equations (\ref{Twb dBEC field good div with P dir evol}) and (\ref{Twb dBEC field good curl with P dir evol}).

Equation (\ref{Twb dBEC eq for pol current gen single part with NO Integral}) gives the final justification of the model of the dipolar BECs with dipole direction evolution presented in Refs. \cite{Andreev EPJ D 13}, \cite{Andreev TransDipBEC12}, \cite{Andreev RPJ 12}. However the consideration presented in this paper gives rather different physical picture behind these equations. Since we have shown that these equations arise from the exchange part of dipole-dipole interaction of bosons with electric dipole moments. The self-consistent part of the dipole-dipole interaction appears to be equal to zero for bosons in the BEC state. Nevertheless, the form of final equations coincides with the formal application of the self-consistent field approximation.

\section{Conclusion}

In spite the fact that the self-consistent field approximation can not be applied to the BECs of particles with the long-range interaction we find that the results obtained earlier on the way of the formal application of the self-consistent field approximation do not contradict to the correct theory.

This unusual conclusion arises due to twisted behavior of the quantum exchange correlations in systems of bosons in the limit of the extremely low temperatures, when all particles are located in the BEC state. The term corresponding the self-consistent field approximation appears at consideration of bosons located in different quantum states, therefore it does not exist for particles located in the single quantum state. Part of quantum correlations has same fate.

The part of quantum correlation related to interaction of particles existing in a same quantum state survives in the BEC state and splits in the product of corresponding one-particle hydrodynamic functions, similar to the formal application of the self-consistent field approximation.

We have this picture in the Euler and the polarisation current evolution equations. Hence our conclusion os correct for both cases of aligned dipoles and dipoles with the dipole direction evolution.

\end{document}